\documentclass[preprint2]{aastex}

\newcommand{\Msol}{\ensuremath{M_{\odot}}}
\newcommand{\Lsol}{\ensuremath{L_{\odot}}}

\shorttitle{Andromeda~XXV}
\shortauthors{Cusano et al.}
\usepackage{color}

\begin{document}

\title{VARIABLE STARS AND STELLAR POPULATIONS IN ANDROMEDA~XXV: III. A CENTRAL CLUSTER OR THE GALAXY NUCLEUS? \altaffilmark{*}}

\author{ FELICE CUSANO\altaffilmark{1}, ALESSIA GAROFALO\altaffilmark{1,2}, GISELLA CLEMENTINI\altaffilmark{1},
MICHELE CIGNONI\altaffilmark{3}, LUCIANA FEDERICI\altaffilmark{1}, 
MARCELLA MARCONI\altaffilmark{4}, VINCENZO RIPEPI\altaffilmark{4}, ILARIA MUSELLA\altaffilmark{4}, VINCENZO TESTA\altaffilmark{5}, 
ROBERTA CARINI\altaffilmark{5}, MARCO FACCINI\altaffilmark{5}}

\affil{$^1$INAF- Osservatorio Astronomico di Bologna, Via Ranzani 1, I - 40127 Bologna, Italy}
\email{felice.cusano@oabo.inaf.it}
\affil{$^2$ Dipartimento di Fisica e Astronomia, Universit\`a di Bologna,  viale Berti Pichat, 6/2, I - 40127 Bologna, Italy}
\affil{$^3$ Dipartimento di Fisica, Universit\`a di Pisa, Largo Bruno Pontecorvo, 3, 56127 Pisa PI}
\email{}
\affil{$^4$INAF- Osservatorio Astronomico di Capodimonte, Salita Moiariello 16, 
I - 80131 Napoli, Italy}
\email{}
\affil{$^5$INAF- Osservatorio Astronomico di Roma, Via di Frascati 33
00040 Monte Porzio Catone, Italy}
\email{}

\altaffiltext{*}{Based on data collected  with the Large Binocular Cameras at the Large Binocular Telescope}

\begin{abstract}
\noindent We present $B$ and $V$ time-series photometry of Andromeda~XXV, the third galaxy in  our program on the Andromeda's satellites, 
 that we have
imaged  with the Large Binocular Cameras of the Large Binocular Telescope. The
field of Andromeda~XXV is found to contain 63 variable stars, for
which we present light curves and characteristics of the light
variation (period, amplitudes, variability type, mean magnitudes,
etc.).  The sample includes 58 RR Lyrae variables (46 fundamental-mode
$-$ RRab, and 12 first-overtone $-$RRc, pulsators), three anomalous
Cepheids, one eclipsing binary system
and one unclassified variable.  The average period
of the RRab stars ($\langle Pab \rangle$ = 0.60 $\sigma=0.04$ days)
and the period-amplitude diagram place Andromeda~XXV in the class of
the Oosterhoff-Intermediate objects.  From the average luminosity of
the RR Lyrae stars we derive for the galaxy a distance modulus of
(m-M)$_0$=$24.63\pm0.17$ mag.
The color-magnitude diagram reveals the presence in Andromeda~XXV of a
single, metal-poor ([Fe/H]=$-$1.8 dex) stellar population as old as
$\sim$ 10-12 Gyr traced by a conspicuous red giant branch and the large
population of RR Lyrae stars.  We discovered a spherically-shaped high
density of stars near the galaxy center. This structure appears to be
at a distance consistent with Andromeda~XXV and we suggest it could
either be a star cluster or the nucleus of Andromeda~XXV. We provide a
summary and compare number and characteristics of the pulsating stars
in the M31 satellites analyzed so far for variability.
\end{abstract}


\section{INTRODUCTION}\label{sec:intro}

In the framework of the $\Lambda$ Cold Dark Matter ($\Lambda$CDM) theory, large galaxies like Andromeda (M31) are formed through
accretion and merging of smaller structures \citep[e.g.][]{bul05}. The satellites
that we observe today around M31 could be the surviving left-overs of the galaxy hierarchical accretion process. 
Their stellar contents can thus provide important hints to reconstruct the star formation history and
the merging episode that led to the formation of M31. The analysis of the color-magnitude diagram (CMD)
represents the most powerful tool for such investigations. However, given the distance of the M31 complex,
ground-based observations barely reach the bottom  of the red giant branch (RGB) of these satellite galaxies and can 
extend down to the horizontal-branch (HB) only when 8-10 m class telescopes are used. 
For this reason pulsating variable stars such as the RR Lyrae stars, which populate the HB of an old ($t \gtrsim$ 10 Gyr) stellar population,
and the brighter and younger Cepheids, 
are powerful  alternative  tools to
trace and characterize the different stellar generations in these systems.\\

This is the third  paper in our series on the M31 satellites based on 
$B$ and $V$ time-series photometry obtained with the Large Binocular Cameras (LBC) of  the Large Binocular Telescope (LBT). 
We characterize 
the resolved stellar populations of the M31 companions using the CMD and the properties   
of variable stars and from this derive hints on 
the nature, origin and fate of Andromeda's  satellites in 
the global context of merging and accretion episodes occurring in  M31.
Details on the survey and results from the study of  Andromeda~XIX (And~XIX) and Andromeda~XXI (And~XXI)
were presented in \citet[][Paper~I]{cus2013} and \citet[][Paper~II]{cus2015}, respectively. In this paper we  
report  results on  Andromeda~XXV (And~XXV).

And~XXV  was  discovered by \citet{rich11} in the context of the PAndAS survey \citep[][and reference therein]{mart13} and later found to be 
a member of the thin plane of satellites identified in M31 by \citet{iba2013}.
 The discovery paper reports a distance modulus of (m-M)$_0$= $24.55\pm0.12$ mag for And~XXV from the luminosity 
of the galaxy HB and a half-light radius (r$_h$) of 3.0$'$ (corresponding to $r_h = 732 \pm 60$ pc at the distance of And~XXV).
Later, \citet{con12}, adopting a bayesian approach to locate the tip of  the red giant branch,   have revised 
the  distance modulus of And~XXV to (m-M)$_0$= $24.33^{+0.07}_{-0.21}$ mag. This is  more than 1 $\sigma$ shorter
than  \citet{rich11}'s 
distance based on the HB luminosity. 
And~XXV has a metallicity of [Fe/H]=$-1.9\pm0.1$ dex (\citealt{col13}) as estimated from the 
Calcium triplet (CaII) of the galaxy red giants. The same authors
measured a velocity dispersion  of $\sigma=3.0^{+1.2}_{-1.1}$ kms$^{-1}$  from 25 spectroscopically confirmed
members. This value is rather low, when compared with the large extension of And~XXV.
\citet{col13} also estimate a mass-to-light ratio inside one r$_h$ of $[M/L]_{r_h}=10.3$ $\Msol / \Lsol$ that, according to the authors, is  consistent with a stellar population
without a dark matter component. In a more recent paper \citet{col14} using as an argument the circular velocity within the galaxy half-light
radius concluded that the mass of And~XXV should have been much more prominent in the past, 
 for the galaxy to be able to form stars. 
On the other hand, the low velocity dispersion of And~XXV's member stars  would naturally arise in the context of the MOND theory
 \citep{mcgaugh2013}.

The paper is organized as follows: observations, data reduction and calibration of And~XXV photometry are presented in Section 2. 
Results on the identification and characterization of the variable stars, 
the catalog of light curves and the Oosterhoff classification of And~XXV are discussed in  Section 3. 
The distance to And~XXV derived from the  RR Lyrae stars is presented in Section 4.
The galaxy CMD along with the spatial distribution of the various stellar components are discussed in Section 5. Section 6 examines the discovery 
of a spherically-shaped high density of stars close to the galaxy center.
Main results of the present study are summarized in Section 7.

\begin{table*}
\begin{center}
\caption[]{Log of And~XXV  observations}
\label{t:obs}
\begin{tabular}{l c c c c }
\hline
\hline
\noalign{\smallskip}
   Dates                 & {\rm Filter}  & N   & Exposure time &  {\rm Seeing (FWHM)}    \\
 	                 &		 &     &	      (s)       &    {\rm (arcsec)}\\
\noalign{\smallskip}
\hline
\noalign{\smallskip}
  October   18, 2011     &   $B$      & 1    & 400    &  1\\
  October 20-24, 2011  &   $B$      & 84     & 400    & 0.8-1  \\ 
                         &             &   &        &        \\  
  October   18, 2011   &   $V$      &  3    &   400  &  1 \\
  October  20-24, 2011   &   $V$      & 84  &  400  & 0.8-1  \\
\hline
\end{tabular}
\end{center}
\normalsize
\end{table*}

\section{OBSERVATIONS AND DATA REDUCTION}
A total of  85 $B$ and 87 $V$ images each of  400 $s$ exposure were obtained
 with the LBC of a region of 23$^{\prime} \times 23^{\prime}$ centered on And~XXV  (R.A.=$00^{h}30^{m}08.9^{s}$,
decl. =$+46^{\circ}51^{\shortmid}07^{\shortparallel}$, J2000.0; \citealt{rich11})  from the 18th to the 24th of October  2011. 
Observations in the $B$ band were acquired with the Blue camera and simultaneous $V$ imaging with the Red camera of the LBC. All images were obtained under favorable conditions of seeing $\le1$ \arcsec.
The log of the observations of And~XXV is provided in Table ~\ref{t:obs}. 

Photometric reduction of And~XXV imaging was carried out following the same procedure as described in detail in Paper~I (And~XIX)  and II (And~XXI).
The PSF photometry was performed using the \texttt{DAOPHOT - ALLSTAR - ALLFRAME} packages \citep{ste87,ste94}.
The Landolt standard fields L92 and SA113,  observed during the run, were  
used to derive calibration equations and tie  the \texttt{DAOPHOT} magnitudes to the Johnson standard system. The new calibration equations\footnote{
$B-b=27.696-0.113\times(b-v)$ r.m.s=0.03,\\
$V-v=27.542-0.060\times(b-v)$ r.m.s=0.03}
are totally consistent with those
derived in Paper~I, once differences in air-mass and exposure times are accounted for.

\section{VARIABLE STARS}
\begin{figure*}[!]
\centering
\includegraphics[width=14cm,height=16cm]{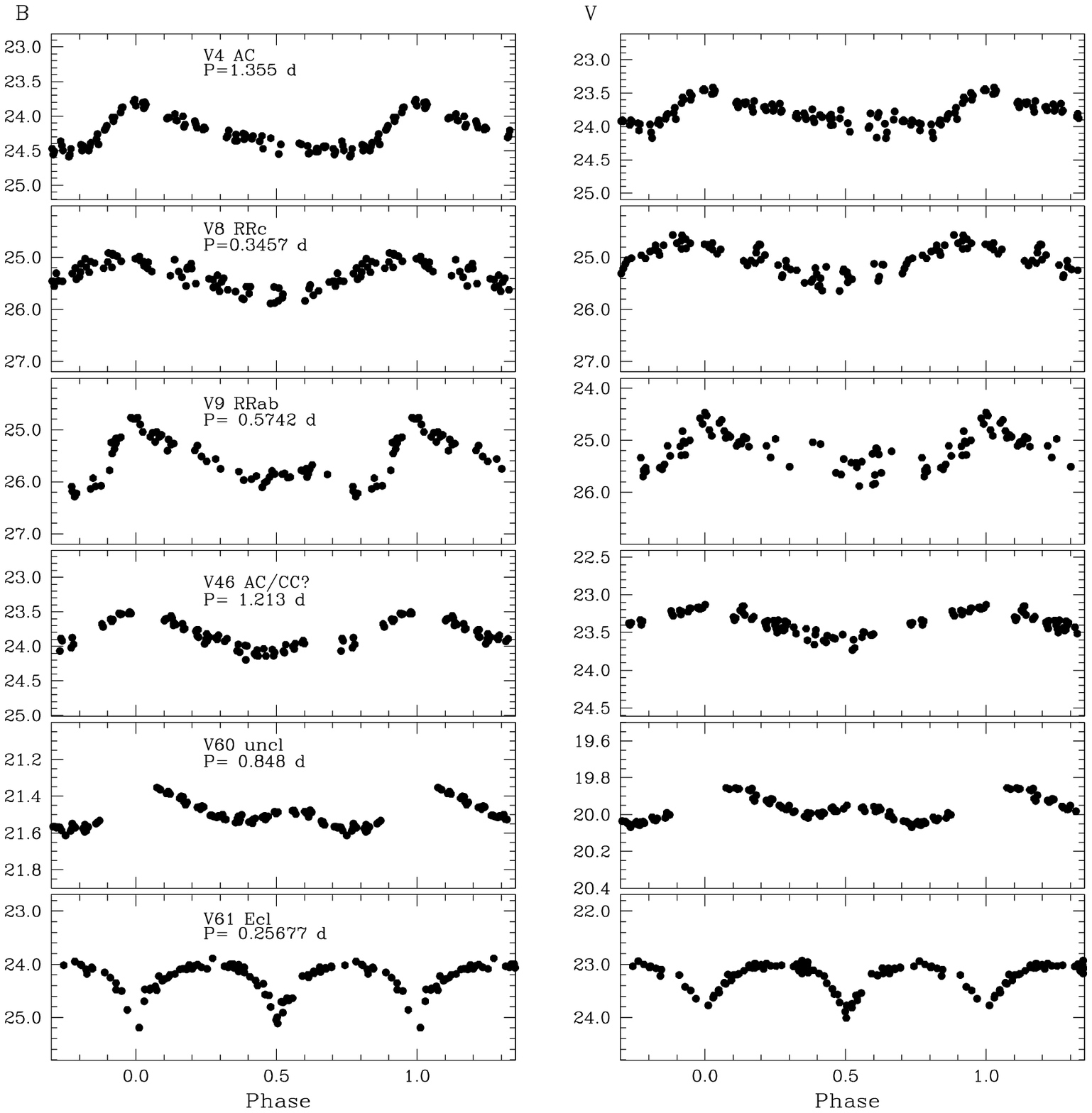}
\caption{Examples of $B$-band (left panels) and $V$-band (right panels)  light curves for different types of variable stars identified in And~XXV. 
Typical internal errors of the single-epoch data range from 0.01 at $B\sim$ 21.30 mag 
(corresponding to the maximum light of the unclassified variable  V60) to 0.30 mag at $B\sim$ 26.20 mag 
(corresponding to the minimum light of the fundamental mode RR Lyrae  V9), and similarly,  from 0.01 mag at $V\sim$ 19.90 mag  to 0.30 mag at $V\sim$ 26.00 mag.}
\label{fig:lca-examples}
\end{figure*}

The search for variable stars  was carried out
starting from the variability index computed in \texttt{DAOMASTER} \citep{ste94}.
The candidate variables were then analyzed by studying  their $B$- and $V$-band  light curves  with  
 the Graphical Analyzer of Time Series (GRATIS), private software developed at the Bologna Observatory 
by P. Montegriffo \citep[see, e.g.,][]{clm00}. More details on this procedure can be found in Paper~I.
We identified, in the $B$, $V$  datasets of And~XXV  a total of 63 variable stars: 
58 RR Lyrae stars (see Section~\ref{sec:rrli}), 3 anomalous Cepheids (AC; see Section~\ref{sec:ac}), 1 eclipsing binary (ECL)
and 1 unclassified variable. 
The properties of the confirmed variable stars in And~XXV are summarized in Table ~\ref{t:1}. 
We ordered the variables with an increasing number based on the proximity to the galaxy center, adopting the coordinates by \citet{rich11}. 
Column 1 lists the star identifier, Columns 2 and 3 give the right ascension and declination (J2000 epoch), respectively, 
obtained from our astrometrized catalogs. Column 4 provides the type of variability. A question mark identifies stars whose classification is uncertain.
Columns 5 and 6 list the period and the Heliocentric Julian Day of maximum light, respectively.  
Columns 7 and 8 give the intensity-weighted mean $B$ and $V$ magnitudes, while Columns 9 and 10 list the corresponding amplitudes of the light variation.
The two solid lines in the table separate the variable stars within the area enclosed once and twice the galaxy  r$_h$.
Example light curves for RR Lyrae and other types of variables in And~XXV are shown in Figure~\ref{fig:lca-examples}. 
The full catalog of light curves is available in the electronic version only.

\begin{table*}
\caption[]{Identification and properties of the variable stars detected in And~XXV. Solid lines separate variable stars located within once and twice the galaxy half-light radius.
}
\scriptsize
\label{t:1}
\begin{tabular}{l c c l l c c c c c }
\hline
\hline
\noalign{\smallskip}
 Name & $\alpha$            &$\delta$       & Type & ~~~P        & Epoch (max)& $\langle B \rangle$ & $\langle V \rangle$ & A$_{B}$ & A$_{V}$ \\ 
 	    &	(2000)   & (2000)   &	         &~(days)& HJD ($-$2455000) & (mag)                       & (mag)         & (mag)     &  (mag)     \\
 	    \noalign{\smallskip}
	    \hline
	    \noalign{\smallskip}
	    
V1  &  00:30:10.688    & +46:51:21.71 &  RRab    & 0.6078   & 852.561 & 25.90    &  25.42  & 1.10    & 0.86 \\
V2  &  00:30:09.937    & +46:51:34.93 &  RRab    & 0.5459   & 851.356 & 25.77    &  25.38  & 1.42    & 0.87  \\
V3  &  00:30:09.306    & +46:51:43.73 &  RRab    & 0.6147   & 852.048 &25.80     & 25.43   &1.42    &  0.86  \\
V4  &  00:30:12.324    & +46:50:58.88 &  AC	  & 1.355    & 854.800 & 24.23    &  23.79  & 0.68   & 0.56\\
V5  &  00:30:05.194    & +46:50:48.71 &  RRc    & 0.3858    & 851.920& 25.82   &  25.41  &  0.72  & 0.36 \\
V6  &  00:30:11.050    & +46:51:58.73 &  RRab	  & 0.609    & 854.732 & 25.58    &  25.23  & 0.98    & 0.76\\
V7  &  00:30:05.122    & +46:50:27.28 &  RRab	  & 0.6116   & 857.708 & 25.80    &  25.39  & 1.00   & 0.72\\
V8  &  00:30:05.122    & +46:50:27.28 & RRc     &  0.3457  & 851.164 & 25.53    &  25.17  & 0.73    & 0.70  \\
V9  &  00:30:14.345    & +46:50:47.45 &  RRab	  & 0.5742   &  856.642& 25.57    &  25.15  & 1.41    & 1.10  \\
V10  &  00:30:04.745    & +46:50:06.57 &  RRab	  & 0.72     &  854.705  & 25.71    &  25.32  & 0.76   & 0.49 \\
V11  &  00:30:14.186    & +46:50:17.44 &  RRab	  & 0.5963   &  856.705 & 25.68    &  25.33  & 1.06   & 0.98\\
V12  &  00:30:12.939    & +46:52:22.79 &  RRab	  & 0.5466   & 854.795 & 25.73    &  25.39  & 1.44   & 1.10\\
V13  & 00:30:02.642    & +46:50:12.32  &  RRab   & 0.7254  & 852.133 & 25.80    &   25.38 &  1.00   &  0.78  \\
V14  &  00:30:02.612    & +46:52:23.61 &  RRab	  & 0.576    & 854.804 & 25.72    &  25.33  & 1.36   & 1.06\\
V15  &  00:30:14.464    & +46:49:28.74 &  RRab	  & 0.591    &  856.663& 25.63    &  25.20  & 1.01   & 0.79\\
V16  & 00:30:17.886     & +46:51:58.24 & RRab   & 0.610    & 851.573  &25.95     & 25.53  &   0.90  &  0.56 \\
V17 &  00:30:18.136    & +46:50:05.58 &  RRab	  & 0.5675   & 855.784 & 25.68    &  25.31  & 1.36   & 1.18\\
V18 &  00:30:02.457    & +46:48:57.10 &  RRc	  & 0.375    &  854.810 & 25.64    &  25.31  & 0.74   & 0.33\\
V19 &  00:30:13.024    & +46:48:31.33 &  RRab	  & 0.577    & 855.587 & 25.66    &  25.28  & 1.22   & 0.95\\
\hline
V20 &  00:29:59.216    & +46:52:40.02 &  RRc	  & 0.3085   & 852.695 & 25.43    &  25.16  & 0.53   & 0.36\\
V21 &  00:30:21.242    & +46:51:02.77 &  RRab	  & 0.562    & 854.804 & 25.41    &  25.07  & 1.60   & 1.25 \\
V22 &  00:29:58.666    & +46:53:04.62 &  RRab   & 0.655   & 857.894 & 25.69    &  25.31  & 0.71   & 0.65\\
V23 &  00:30:21.905    & +46:51:36.72 &  RRc	  & 0.3668   & 858.900 & 25.49    &  25.15  & 0.81   & 0.61\\
V24 &  00:30:18.610    & +46:53:37.02 &  RRab	  & 0.537    & 854.816 & 25.66    &  25.29  & 1.89   & 1.28\\
V25 &  00:30:15.060    & +46:47:56.28 &  RRab	  & 0.637    & 856.670 & 25.73    &  25.27  & 0.80   & 0.51\\
V26 &  00:30:23.121    & +46:51:10.67 &  RRc	  & 0.405    & 856.782 & 25.61    &  25.26  & 0.70   & 0.47\\
V27 &  00:30:21.141    & +46:49:12.98 &  RRab	  & 0.547    & 857.652 & 25.60    &  25.22  & 1.25   & 0.97\\
V28 &  00:30:17.693    & +46:48:10.12 &  RRab	  & 0.565    &  856.747 & 25.62    &  25.25  & 1.42   & 1.12\\
V29 &  00:29:55.136    & +46:49:28.51 &  RRab	  & 0.632    & 856.860 & 25.73    &  25.26  & 0.781   & 0.56\\
V30 &  00:30:03.691    & +46:47:25.31 &  RRc	  & 0.368    & 856.767 & 25.55    &  25.20  & 0.452   & 0.41\\
V31 &  00:30:13.107    & +46:47:15.13 &  RRab	  & 0.669    & 856.780 & 25.50    &  24.88  & 0.845   & 0.51\\
V32 &  00:30:11.748    & +46:55:04.50 &  RRc	  & 0.378    & 856.677 & 25.60    &  25.34  & 0.66    & 0.59\\
V33 &  00:29:58.771    & +46:47:48.94 &  RRab	  & 0.577    &  856.850 & 25.61    &  25.24  & 0.87   & 0.72\\
V34 &  00:30:26.025    & +46:51:13.58 &  RRab	  & 0.5575   &  858.894& 25.65    &  25.30  & 1.16    & 0.81 \\
V35 &  00:30:19.772    & +46:47:45.21 &  RRab	  & 0.579    &  854.780& 25.64    &  25.17  & 1.50   & 1.17\\
V36 &  00:30:26.764    & +46:50:36.83 &  RRab	  & 0.5678   &  858.615& 25.47    &  25.14  & 1.71   & 1.32\\
V37 &  00:30:02.239    & +46:46:45.76 &  RRab	  & 0.6992   & 857.747 & 25.64    &  25.23  & 0.80   & 0.62\\
V38 &  00:30:18.484    & +46:47:02.00 &  RRab	  & 0.67     &  856.802 & 25.63    &  25.30  & 0.72    & 0.47\\
V39 &  00:30:22.544    & +46:54:27.27 &  RRab	  & 0.5766   & 858.601 & 25.54    &  25.22  & 1.21   & 1.00\\
V40 &  00:30:18.502    & +46:46:50.06 &  RRab	  & 0.621    & 858.638 & 25.66    &  25.27  & 1.37   & 0.82\\
V41 &  00:29:55.859    & +46:47:19.20 &  RRab	  & 0.618    & 856.664 & 25.66    &  25.29  & 0.82   & 0.67\\
V43 &  00:29:48.512    & +46:51:50.74 &  RRc	  & 0.37     & 855.660 & 25.59    &  25.17  & 0.64   & 0.34 \\
\hline
V44 &  00:29:48.589    & +46:49:46.67 &  RRab	  & 0.615    &  858.598& 25.59    &  25.25  & 0.92   & 0.69\\
V45 &  00:30:04.316    & +46:44:45.19 &  RRab	  & 0.5867   &   855.842& 25.59    &  25.33  & 1.13   & 0.93 \\
V46 &  00:29:42.803    & +46:51:51.83 &  AC/CC?     &  1.213  & 852.697  & 23.82    &   23.38 &  0.62  &  0.40    \\
V47 &  00:30:08.635    & +46:44:29.81 &  RRab	  & 0.6322   & 854.715  & 25.65    &  25.36  & 1.02   & 0.81\\
V48 &  00:29:57.962    & +46:44:56.31 &  RRab	  & 0.6325   & 857.792  & 25.74    &  25.36  & 0.81   & 0.70\\
V49 &  00:30:25.549    & +46:56:26.38 &  RRab     &  0.592   & 855.790  & 25.65    &  25.33  & 1.05    & 0.92  \\
V50 &  00:29:41.661    & +46:50:24.07 &  RRab     &  0.579   & 856.863  & 25.65    &  25.45  & 0.81    & 0.67   \\
V51 &  00:30:17.641    & +46:57:49.18 &  RRab     & 0.6371   & 855.593  & 25.43    &  25.08  & 1.13    &  0.93   \\
V52 &  00:30:09.862    & +46:43:59.74 &  RRab	  & 0.6243   & 856.652 & 25.58    &  25.20  & 0.97   & 0.71\\
V53 &  00:30:20.789    & +46:44:32.70 &  RRc?	  & 0.3934   & 854.826 & 25.54    &  25.26  & 0.56   & 0.42\\
V54 &  00:30:24.470    & +46:57:13.95 &  RRab     &   0.635  & 855.580 & 25.66    &  25.20  & 0.80   &  0.65  \\
V55 &  00:30:37.347    & +46:52:57.46 &  RRab     & 0.681    & 851.572 & 25.61    & 25.231  & 0.80   & 0.66    \\
V56 &  00:29:50.346    & +46:44:58.80 &  RRc	  & 0.299    &  858.907 & 25.66    &  25.39  & 0.67   & 0.53\\
V57 &  00:30:16.060    & +46:42:25.70 &  RRab	  & 0.57     & 856.764 & 25.77    &  25.38  & 0.97   & 0.72\\
V58 &  00:29:32.588    & +46:49:37.73 &  AC       & 0.531    &  855.810& 24.53   & 24.24    & 0.90   & 0.76   \\
V59 &  00:29:56.648    & +46:42:10.13 &  RRab	  & 0.642    & 856.847 & 25.45    &  25.22  & 1.201   & 1.087\\
V60* &  00:30:13.652    & +46:41:30.72 &  uncl	  & 0.848    & 854.595 & 21.48    &  19.96  & 0.239   & 0.197\\
V61* &  00:30:06.316    & +46:41:26.44 &  ECL    & 0.25677  & 856.760 & 24.24    &  23.22  & 1.39    & 0.886\\
V62 &  00:30:14.030    & +46:41:22.26 &  RRab	  & 0.5337   & 856.732 & 25.67    &  25.22  & 1.58    & 1.233\\
V63* &  00:29:49.898    & +46:49:07.68 &  RRab	  & 0.5486   & 855.629  & 25.55    &  25.16  & 1.483   & 1.157\\
\hline 
*Field variable stars\\
\end{tabular}
\normalsize
\end{table*}

\subsection{RR LYRAE STARS}\label{sec:rrli}

\begin{figure}[t!]
\centering
\includegraphics[width=8.0cm,clip]{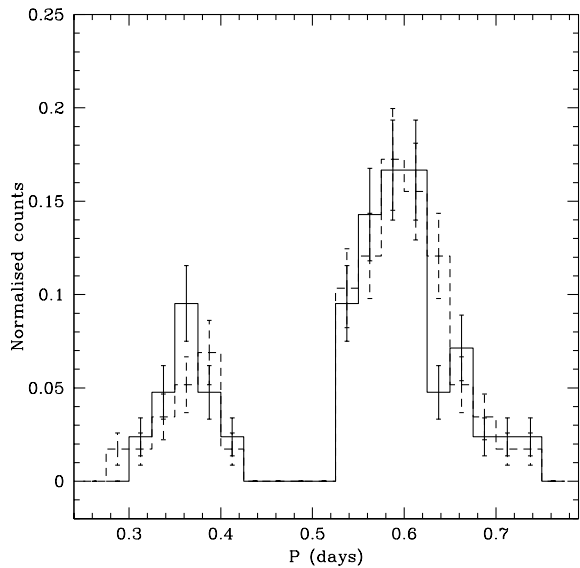}
\caption{The dashed histogram  represents the distribution of the periods of all the RR Lyrae stars identified in And~XXV.
 The continuous line histogram is for RR Lyrae enclosed in twice the galaxy r$_h$. 
The bin size is 0.025 days.}
\label{fig:hist}
\end{figure}

In the field of And~XXV we have identified a total of 58 RR Lyrae stars:  46 fundamental mode (RRab) and 12 first-overtone (RRc) stars.
Fifty-seven of them likely belong to And~XXV, whereas 
V63 is $\sim12$ arcmin  away from the center of the galaxy and 
is most probably an RR Lyrae of the M31 halo.  We do not include this star when deriving properties of And~XXV such as the distance, the Oosterhoff type, etc. 
Figure~\ref{fig:hist} shows the period distribution of the RR Lyrae stars.
The average period of the whole sample of 45 RRab stars in And~XXV is $\langle$P$_{\rm ab}\rangle$=0.61 d ($\sigma$=0.05 d)
and becomes $\langle$P$_{\rm ab}\rangle$=0.60 d ($\sigma$=0.05 d) if we only consider 
15 RRab's within the r$_h$ convolved with the galaxy ellipticity, while it is $\langle$P$_{\rm ab}\rangle$=0.60 d ($\sigma$=0.05 d) 
for the 32 RRab's within twice the r$_h$.
In any case, And~XXV is classified as an  Oosterhoff-Intermediate (Oo~Int) system (\citealt{oos39,cat09}). 

The fraction of RRc stars in And~XXV over the whole number of RR Lyrae stars is f$_c$= N$_c$/N$_{\rm ab+c}=0.21\pm0.07$, 
that is  intermediate between what expected for Oosterhoff~II   (Oo~II; f$_c\sim0.44$) 
and  Oosterhoff~I (Oo~I; f$_c\sim0.17$) systems \citep{cat09}, again  confirming the classification of And~XXV as Oo~Int.

\begin{figure*}[t!]
\centering
\includegraphics[trim=25 10 0 0 clip, width=0.48\linewidth]{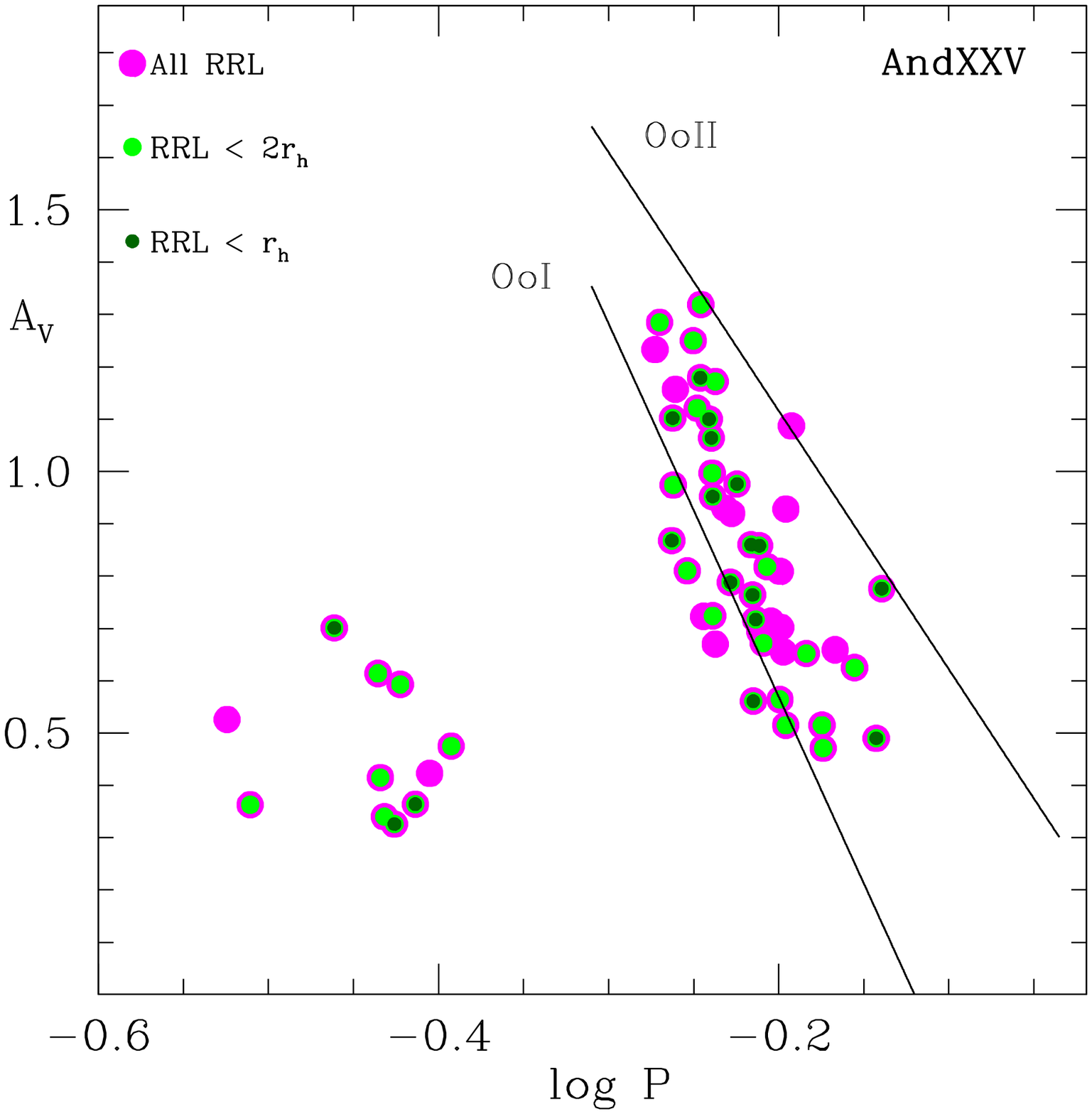}\includegraphics[trim=25 10 0 0 clip, width=0.48\linewidth]{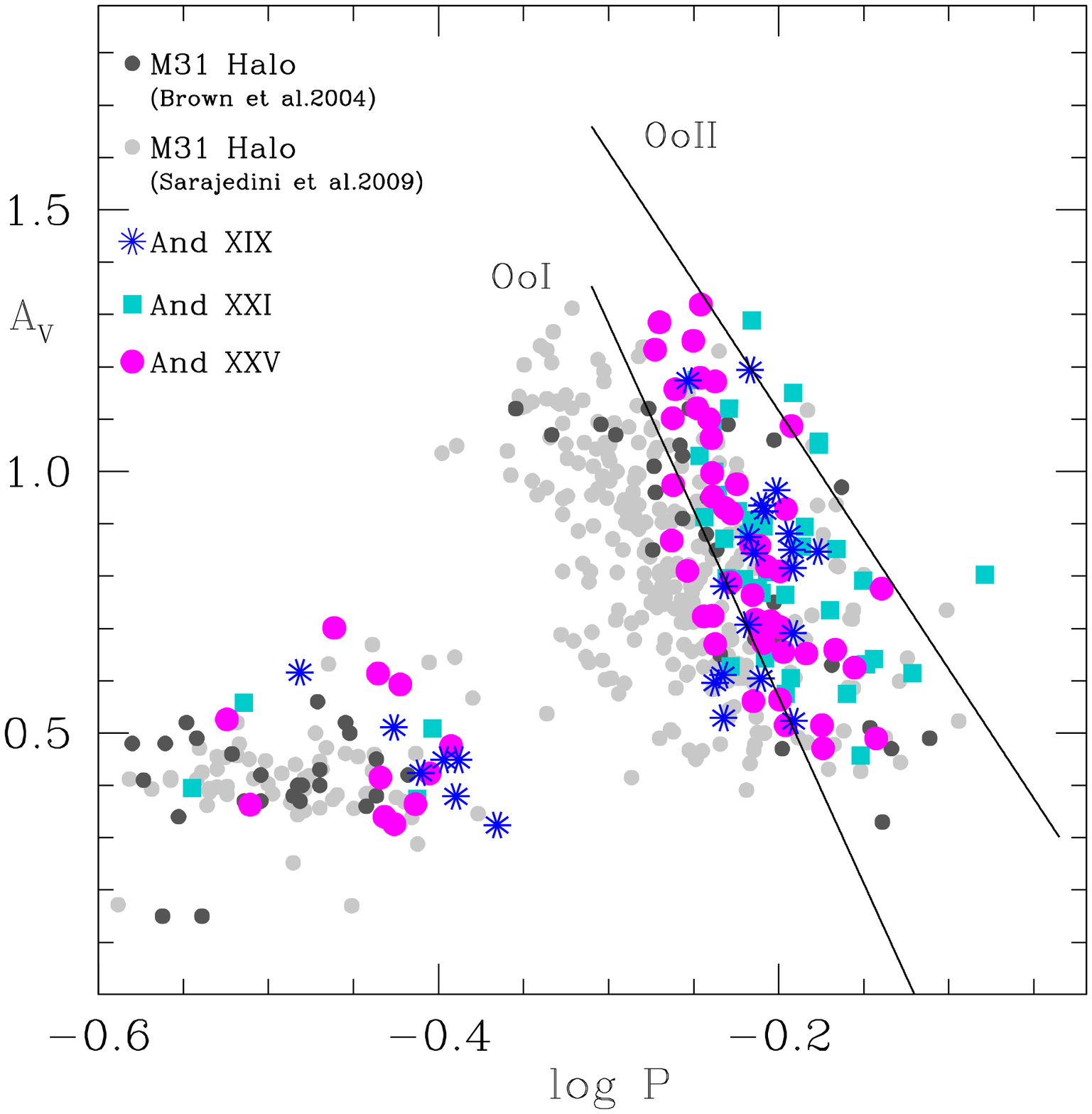}
\caption[]{ $Left$:Period-amplitude diagram of the RR Lyrae stars in And~XXV. The solid lines are the ridge lines of
 the Oo~I and Oo~II RR Lyrae stars (Clement \& Rowe  2000). Dark  and light green circles 
 are the RR Lyrae stars inside once and twice the r$_h$ of the galaxy, respectively. Purple circles are all the RR Lyrae stars. 
 $Right$: Same as  in the left panel combined with the RR Lyrae stars in And~XIX (blue asterisks), And~XXI (cyan squares) and in three HST fields in the M31's halo 
 from \cite{sar09} (grey symbols) and \cite{bro04} (dark-grey symbols), respectively.}
\label{fig:bayl}
\end{figure*}

The period-amplitude diagram (also known as Bailey diagram, \citealt{Bai1902}) of the RR Lyrae stars in And~XXV is shown in the left panel 
of Figure~\ref{fig:bayl}
together with the loci defined by  RR Lyrae stars in the Oo~I Galactic globular cluster M3 (lower line) 
and the Oo~II  globular cluster $\omega$ Cen (upper line),  
according to \citet{cle00}. 
We separated the RR Lyrae stars in three groups depending on the position 
inside once and  twice  the area delimited by the r$_h$ convolved with the galaxy ellipticity, and in the whole LBC field of view (FOV). 
Almost all RR Lyrae stars fall on the Oo~I locus or between the two lines but keeping closer to the Oo~I position. 
Only three  RR Lyrae stars are near the Oo~II locus. 
In the right panel of Figure~\ref{fig:bayl} the period-amplitude
diagram of And~XXV's RR Lyrae stars is compared to the distribution of
RR Lyrae in And~XIX, And~XXI and in three halo fields at 4, 6 kpc
\citep{sar09} and 35 kpc \citep{bro04} from the center of M31,
respectively.  The RR Lyrae stars in the M31 halo conform more to the
Oo~I type, while the  dwarf Spheroidal galaxy (dSph) satellites are more Oo~Int with a slight
trend towards Oo~I type, which decreases with increasing the distance from
the M31 center.

The RR Lyrae stars trace the old populations of a galaxy and from their positions is thus possible to draw a first 
picture of the geometry of And~XXV. 
The spatial distribution of the RR Lyrae stars in
And~XXV is shown in Figure \ref{fig:spa} (filled red circles), and
reveals the elongated structure of this galaxy.  Adopting the
structural parameters derived by \citet{rich11}, many RR Lyrae stars
appear to be placed beyond the galaxy half-light radius, traced by the
inner black ellipse, and create a protrusion of stars that extends beyond twice the $r_h$.  In particular, 
more than half of the RR Lyrae stars (37 out of 57) are located
southward with respect to the galaxy center and suggest a stretched
distribution of And~XXV's stars. Most of them are aligned along the
direction toward M31.
 
\subsection{ANOMALOUS CEPHEIDS}\label{sec:ac}

In And~XXV we identified three variable stars, namely, V4, V46 and V58, that fall inside the instability strip  \citep[see ][]{mar04}, 
but have luminosities exceeding the average magnitude of the RR Lyrae stars by about 1 mag. 
Following the procedure discussed in Papers I and II, 
by the cross study with isochrones in the CMD, 
the period-Wesenheit ($PW$) relations for ACs and classical Cepheids (CCs) 
  and the  analysis of the light curves we classify these three stars  as  ACs.
  However the classification for the star V46 is more uncertain and this star can be also seen as a short period CC.

  To obtain the Wesenheit index of these 3 variables we subtracted the
  color term multiplied by 3.1 \citep[adopting the ][law for extinction]{card1989} to the absolute magnitudes derived by their intensity-averaged apparent visual magnitudes,  $\langle V \rangle $,  
  adopting the distance modulus of  (m-M)$_0$=$24.63\pm0.17$ mag as derived from the RR Lyrae stars (see
  Section~\ref{sec:dist}).  In the left panel of Figure~\ref{fig:plac}
  are displayed the $PW$ relations for ACs in the Large Magellanic
  Cloud (LMC, solid lines) derived by
  \citet{ripe14}\footnote{\citet{ripe14}'s relations were derived for
    the $V$ and $I$ bands, and we have converted them to $B$ and $V$
    using equation 12 of \citet{mar04}.}.   
    V4, V46   and V58  fall within 1$\sigma$ of the
  \citet{ripe14} relations for fundamental mode and first-overtone
  pulsators.  In the right panel of Figure~\ref{fig:plac} we compare
  the three variables with the $PW$ relations of the LMC CCs by \citet{jac16}. Clearly, V4 does not fit the CC $PW$ relations, while both V46 and V58 are within 
 1  $\sigma$ of the fundamental mode CC $PW$. However,  looking at the shape of the light curve 
  V58 shows the typical
  trend variation of an AC (see Paper I and II for examples).
Therefore,  we finally classify  V58 as AC, while for V46 we
give an intermediate classification between AC and CC.  
 Since both 
  V46 and V58 are outside twice the area defined by the galaxy r$_h$,  
 they could either belong to a background/foreground
  feature of the M31 field,  or could  be recently formed stars due to a peripheral episode of star formation in And~XXV
  triggered by tidal interaction. In any case, given the uncertain membership of V46 and V58, we have computed 
 the specific frequency of ACs in And~XXV under the
  assumption that V4 is the only  AC belonging to the galaxy. This is shown in
  Figure~\ref{fig:spec} together with the AC specific frequency in a
  number of MW and M31 dwarf satellites.  And~XXV does not follow the
  relation traced by the other galaxies.  Moreover, to make a more
  consistent comparison between the ACs discovered in this paper, in Papers~I and ~II, and those presented in other studies
  \citep[][and reference therein]{pri05}, we have revised the AC 
  specific frequency in dwarf galaxies analyzed in those previous studies, considering as bona fide ACs only 
  those falling  
  within 1  $\sigma$ of the PW relations for ACs.  This tool has confirmed
  the AC number in the following satellites: Leo~I, Leo~II, Draco,
  Ursa Minor, Carina, Sculptor, and Fornax but has reduced it in And~I and
  And~II (from 1 to 0), And~III (from 5 to 2), And~VI (from 6 to 4)
  and Sextans (from 4 to 2). The red solid lines plotted in both
  panels of Figure~\ref{fig:spec} are the least-square fits obtained
  adopting the revised specific frequencies computed as discussed above. They are described by the following
  equations:
 \begin{equation}
 \rm logS=0.23(\pm 0.57)\times Mv+2.22(\pm 0.57)
 \end{equation}
 \begin{equation}
 \rm logS=-1.39(\pm 0.44)\times [Fe/H]-2.62(\pm 0.83)
 \end{equation}

Black lines mark instead the least-square fits originally obtained in \citet{pri05}. 
  
  Grey dashed arrows indicate the position the revised galaxies would occupy in these plots if we used 
  their  literature AC specific frequencies.  In the right panel  new and 
  previous fits are in perfect agreement, while in the left panel the slope of the 
  new fit is less steep indicating that for fainter galaxies a 
  smaller number of ACs is expected.

\subsection{OTHER VARIABLES}
\begin{figure}[t!]
\centering
\includegraphics[trim=40 15 0 0 clip, width=1.\linewidth]{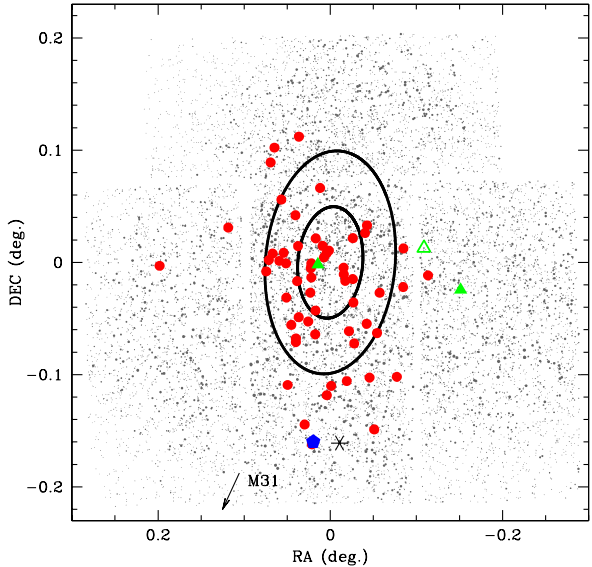}
\caption[]{Spatial distribution of the variable stars discovered in the whole LBT FOV.  
The black ellipses represent the area enclosed respectively by once and twice the r$_h$ of And~XXV defined by Richardson et al. (2011). 
Red filled circles mark the RR Lyrae stars, green filled triangles  the ACs, open green triangle is the uncertain AC/CC, 
the blue pentagon is the unclassified variable and  the black asterisk is the ECL. The arrow points to the  direction of M31.
}
\label{fig:spa}
\end{figure}

\begin{figure}[t!]
\begin{center}
\includegraphics[trim=40 25 0 0 clip, width=1.05\linewidth]{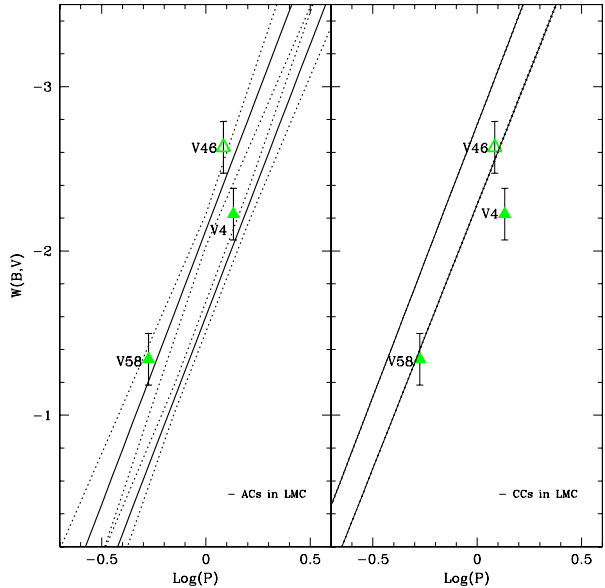}
\end{center}
\caption[]{Position of V4, V46 and V58 with respect to 
 the $PW$ relations for ACs (left panel; Ripepi et al. 2014) and CCs in the LMC  (right panel;  Jacyszyn-Dobrzeniecka et al. 2016). 
 The 95\% confidence contours are shown as dotted lines. For the CCs the errors of the fits are very small and the 
condifence contours are very close to the fits.}
\label{fig:plac}
\end{figure}

\begin{figure*}[]
\centering
\includegraphics*[scale=0.5]{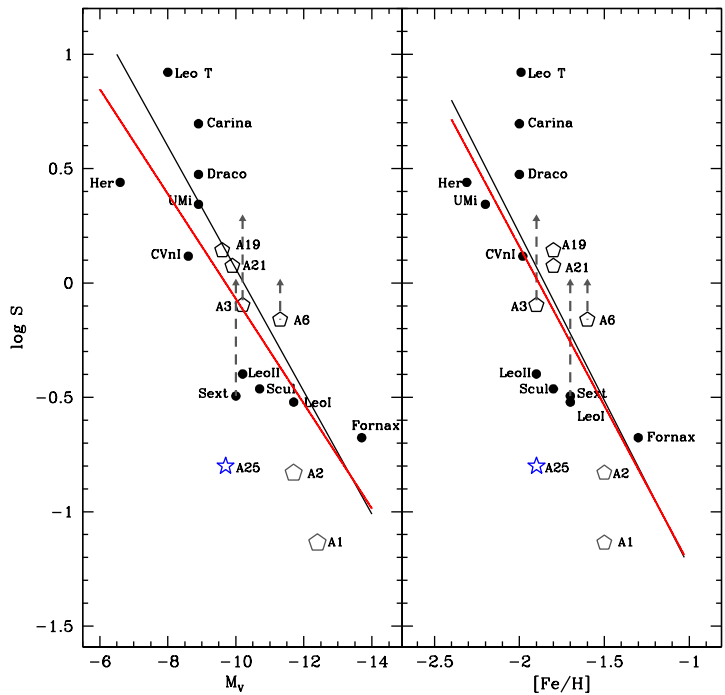}

\caption[]{$Left$: Specific frequency of ACs in dwarf satellites of the MW (filled circles)
and M31 (open pentagons) versus absolute visual magnitude. And~XXV is represented by the blue star. 
Black lines are the least-squares fits of ACs by \citet{pri05} while the red solid lines mark the least-squares fits we have
 been obtaining 
using only  ACs confirmed by the comparison with the PW relations for ACs (see text for details).
$Right$: Same as in the left panel, but versus metallicity of the parent galaxy.}
\label{fig:spec}
\end{figure*}

In the field of And~XXV we discovered two variable stars, namely, V60 and V61, 
that in the CMD (Figure~\ref{fig:cmd}) are outside the instability strip for Cepheids and RR Lyrae stars \citep{mar04}.
V61 was  classified by the light curve analysis as 
an  ECL, while we were unable to give a clear classification for V60.  
The position in the CMD (see Figure~\ref{fig:cmd}) and the projected spatial distribution 
(Figure~\ref{fig:spa}) suggest them to belong more likely to the M31 field or to Milky Way (MW). 
V60 in particular, with a $B-V$ $\sim 1.5$ mag and a $V~20$ mag is a field star, beyond any doubts.

\section{DISTANCE}\label{sec:dist}

The $V$ average magnitude of the whole sample of RR Lyrae stars in And~XXV is  $\langle V(RR) \rangle=25.26\pm0.10$ mag (average on 57 stars).
If we exclude V31 that is  0.39 mag brighter than the RR Lyrae average magnitude
and can be either a blended star or an evolved RR Lyrae we obtain $\langle V(RR) \rangle=25.27\pm0.09$ mag. 
Considering instead the subsample of RR Lyrae  inside an area delimited by And~XXV r$_h$  the average $V$ magnitude  
is $\langle V(RR) \rangle=25.33\pm0.09$ mag (average on 18 stars). 
As for Paper I and II, to correct for interstellar extinction we derived the reddening from the galaxy RR Lyrae stars  
adopting the method described by \citet{pier02}.
For the sample of RR Lyrae within the r$_h$, the reddening is $E(B-V)=0.08\pm0.04$ mag, 
for the whole sample $E(B-V)=0.05\pm0.04$ mag. These values are slightly 
 smaller  than  derived by \citet[][$E(B-V)=0.10\pm0.06$ mag]{sch98}, but still consistent within 1 $\sigma$.
To compute the distance to  And~XXV we adopt the average $V$ magnitude and reddening 
obtained from the RR Lyrae inside  the r$_h$,   M$_{\rm V}=0.54\pm0.09$ mag  
for the absolute visual magnitude of RR Lyrae stars with metallicity of [Fe/H] = $-$1.5 dex \citep{cle03} and
$\frac{\Delta {\rm M_V}}{\Delta {\rm[Fe/H]}}=-0.214\pm0.047$ mag/dex \citep{cle03,gra04} for the 
slope of the RR Lyrae luminosity-metallicity relation. 
For the metallicity of And~XXV we adopt the value  [Fe/H]=$-1.9\pm0.1$ dex as derived spectroscopically by \citet{col13}.
The distance modulus of And~XXV  derived from the RR Lyrae stars  is thus (m-M)$_0$=$24.63\pm0.17$  mag.
This estimate is in very good agreement with the result of
\citet{rich11}, (m-M)$_0$=$24.55\pm0.12$ mag, who used the luminosity
of the HB to infer the galaxy distance, but it is more than 1$\sigma$ longer than obtained by 
\citet{con12}, (m-M)$_0$= $24.33^{+0.07}_{-0.21}$ mag, who used the
luminosity of the RGB tip. We suspect that the distance derived in \citet{con12} might be hampered by the uncertain location of  the RGB tip which is scarcely populated
in And~XXV. 

\section{CMD AND STELLAR POPULATIONS}\label{sec:cmd}
The CMD of And~XXV is shown in Figure~\ref{fig:cmd}, where in the
left panel are plotted only sources located within the area
delimited by the galaxy half-light radius and the ellipticity given by
\citet{rich11}, in the central panel are displayed sources enclosed between once and twice the
$r_{h}$, whereas the right panel shows the CMD of sources in the whole 
LBC FOV. We minimized the presence of background
  galaxies and peculiar sources by selecting only sources with DAOPHOT quality image parameters 
   $-0.35 \le$ Sharpness $\le 0.35$ and $\chi < 1.5 $.  Red filled
  circles are the RR Lyrae stars, green filled triangles are the ACs (the open green triangle is the 
  uncertain AC/CC), 
  the blue pentagon is the unclassified variable, while
  the ECL is marked by a
  black asterisk.  The main features of the CMD are the
  RGB and the HB traced by the RR Lyrae stars.  The majority of stars
  filling up the redder part of the CMD at $B-V\sim 1.5-1.7$ mag and
  going beyond $V=22$ mag are foreground field objects. V60, the unclassified variable
  located in this region of the CMD is most
  probably a field variable star belonging to the MW field.

 The narrow RGB and the poorly  populated  red HB   suggest  that And~XXV hosts a dominant single old stellar population.
In Figure~\ref{fig:pana} the CMD features and the position of the variable stars are compared with  Padua isochrones
obtained using 
the CMD 2.5 web interface ($http://stev.oapd.inaf.it/cgi-bin/cmd$) based on models from \citet{bres12}. 
Isochrones for ages from 9 to 13 Gyr and metallicity Z=0.0003-0.0004 fit well the RGB and the position of the RR Lyrae stars. 
The red HB is also well fitted by the  isochrones in these age and metallicity ranges.

The only AC discovered within twice the $r_{h}$ of And~XXV can either be the result of coalescence in a
binary system as old as the RR Lyrae stars, in which mass transfer acted in the
last Gyr, or an intermediate young (1-2 Gyr old) star with metallicity in the range
Z=0.0001-0.0006 (as widely discussed in Paper~I and Paper~II).  The
paucity of ACs in And~XXV gives indication that there is lack of a
young/intermediate age stellar population in this galaxy, when compared to
And~XIX and And~XXI. We selected stars in the CMD of sources in the whole LBC FOV that have color and magnitude in the  range of ACs
($0.2\le B-V \le0.6, 23.0 \le V \le 24.5 $) and we did not find any
clear distribution/concentration of stars associated with this selection. The star
formation in And~XXV most probably was quenched by some external event
as, for instance,  tidal stripping by large nearby galaxies like the NGC147/NG185 pair (see  next Section) or
from M31 itself. In this scenario, the presence of at least one CC and one AC
  off-centered with respect to And ~XXV could indicate that a mild recent activity has been
  triggered from the outside.
  
\subsection{PROJECTED SPATIAL DISTRIBUTIONS}\label{spatial}

\begin{figure*}[]
\centering
\includegraphics[scale=0.7]{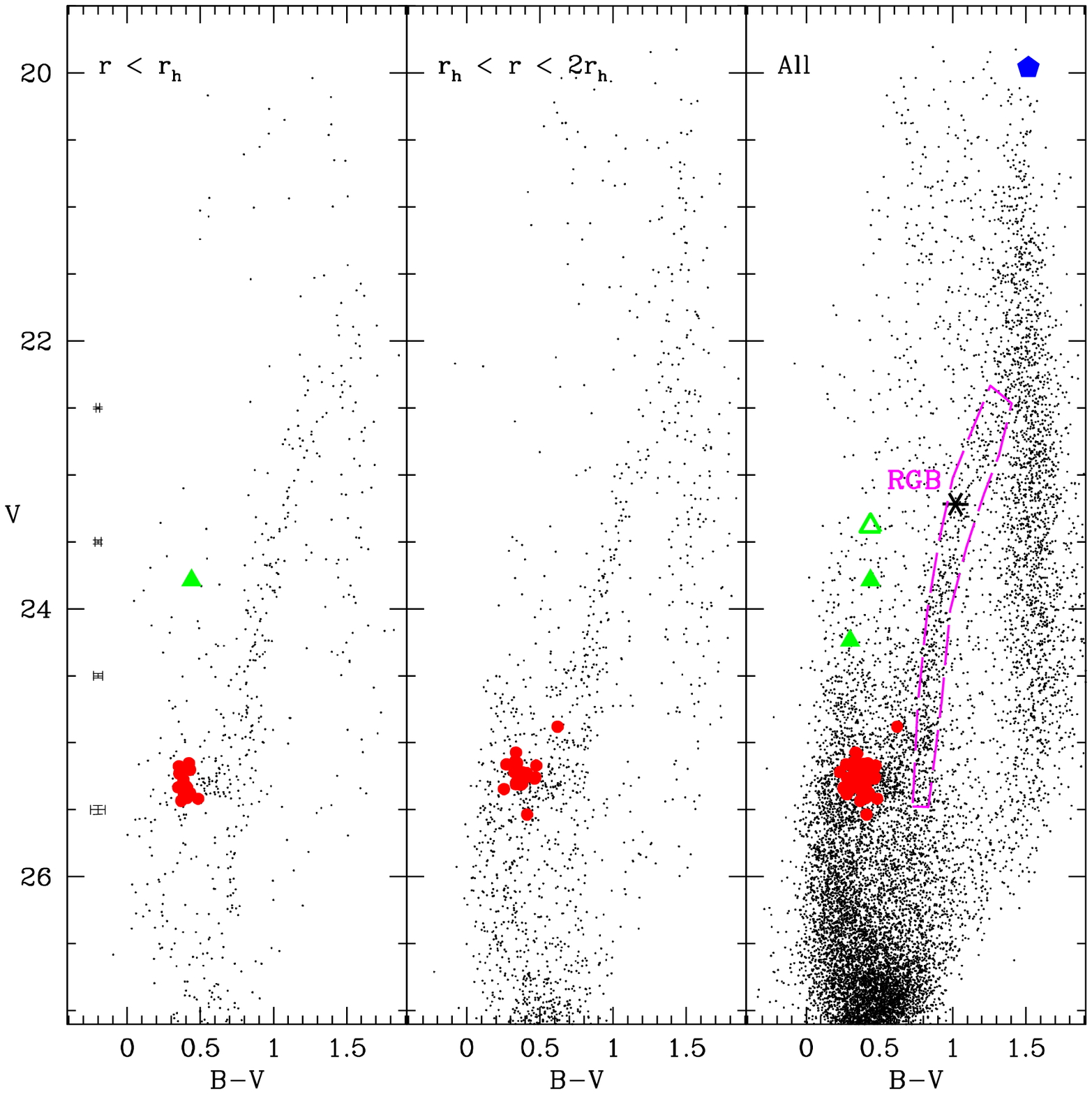}
\caption[]{$Left$: CMD of sources in our photometric catalog with  $-0.35 \le$ Sharpness $\le 0.35$ and with $\chi < 1.5 $ and 
 located within the area delimited by the galaxy half-light radius and the ellipticity  by Richardson et al. (2011). Red  
circles mark the RR Lyrae stars while green triangles are  the ACs. $Center$: Same as in the left panel, but 
within the area between once and twice the galaxy r$_h$. $Right$:
As in the left panel, but considering  sources in the whole LBC FOV. The black asterisk is the ECL, the blue pentagon is the unclassified variable,
while the open green triangle is the uncertain AC/CC. }
\label{fig:cmd}
\end{figure*}

\begin{figure*}[]
\centering
\includegraphics[width=13cm,height=16.5cm]{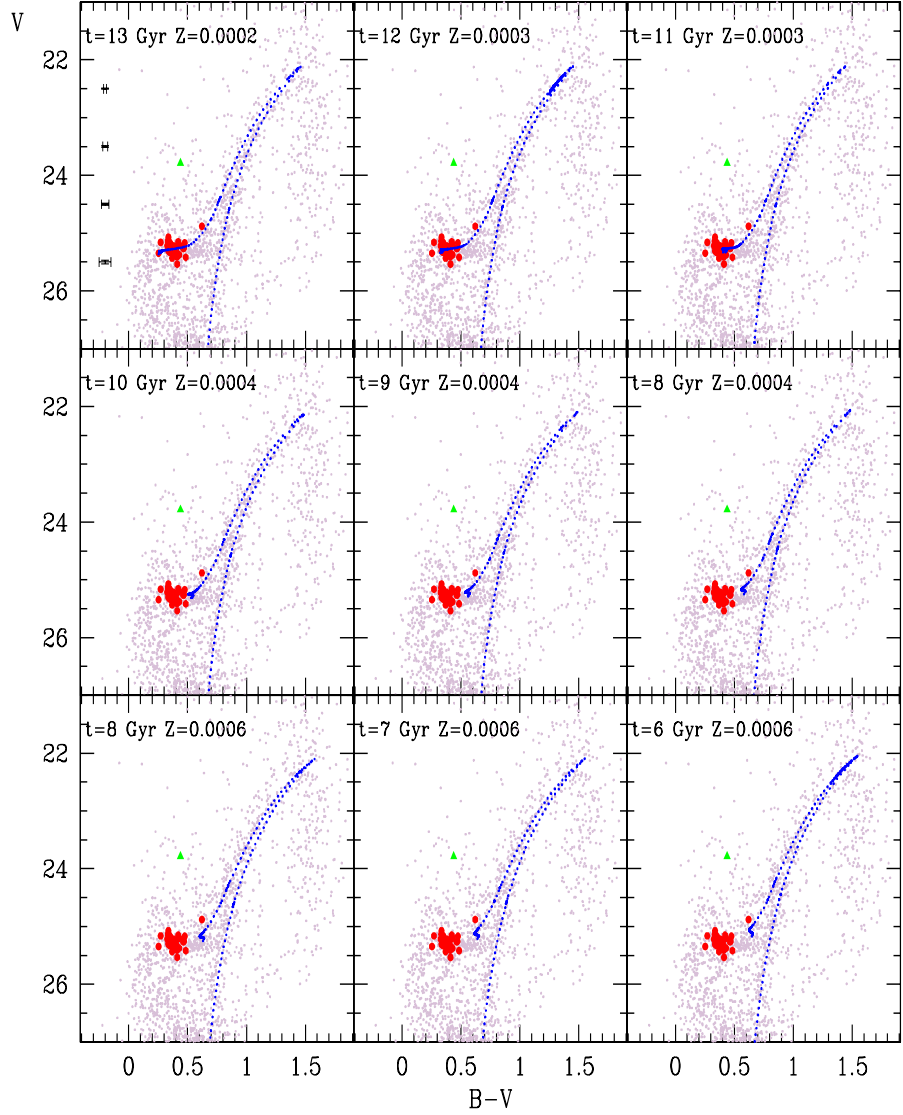}
\caption[]{CMD of sources contained within twice the r$_h$ of And~XXV, overlaid by Padua stellar isochrones (Bressan et al. 2012) 
with different age (from t=13 Gyr to 6 Gyr) and metallicity (Z=0.0003, Z=0.0004 and Z=0.0006; from top to bottom).  
Red filled circles are RR Lyrae stars,  the green filled triangle is the AC (V4).}
\label{fig:pana}
\end{figure*}

\begin{figure*}[t!]
\centering
\includegraphics*[trim=25 10 0 0 clip, width=0.48\linewidth]{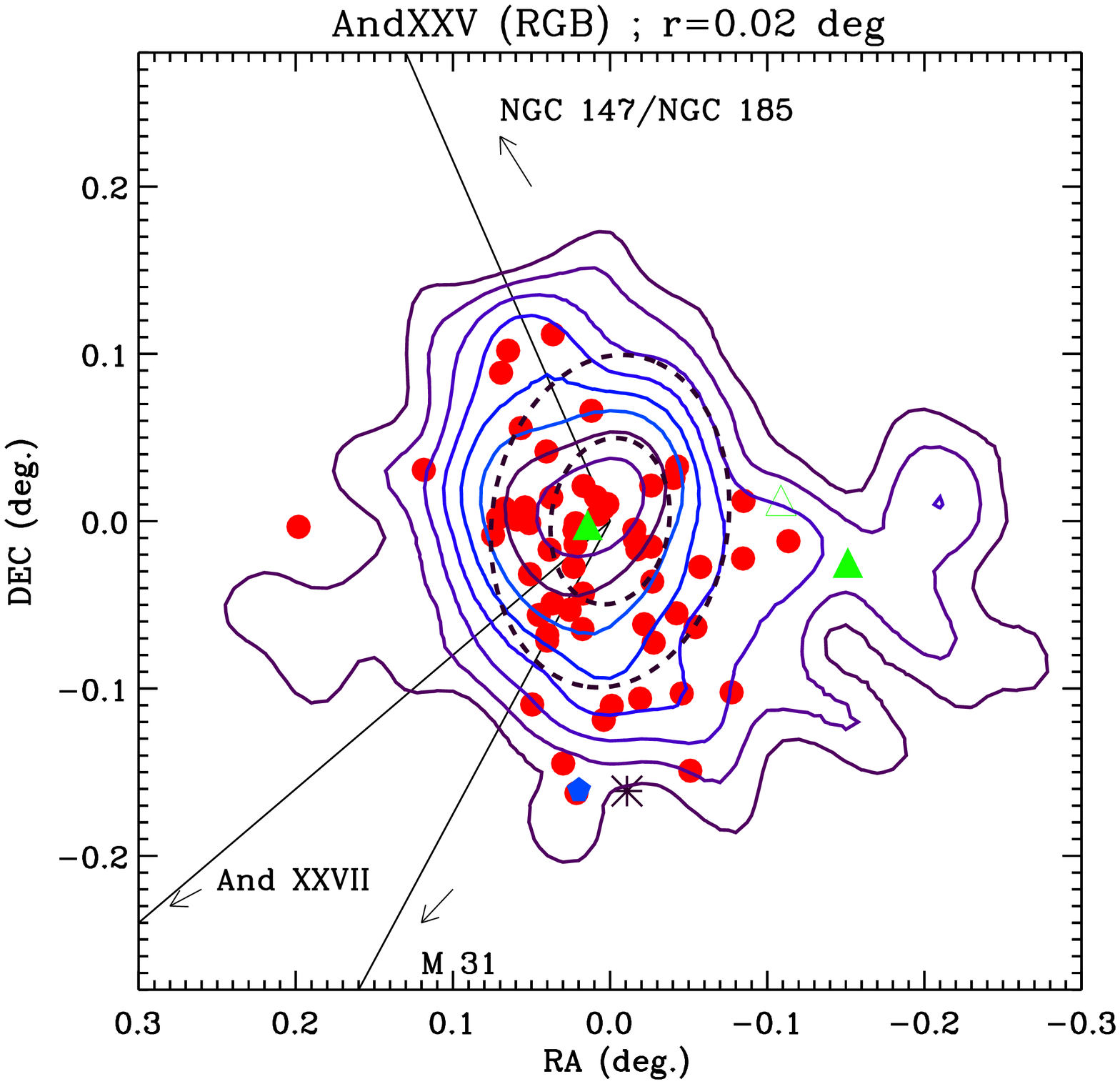}\includegraphics*[trim=25 10 0 0 clip, width=0.48\linewidth]{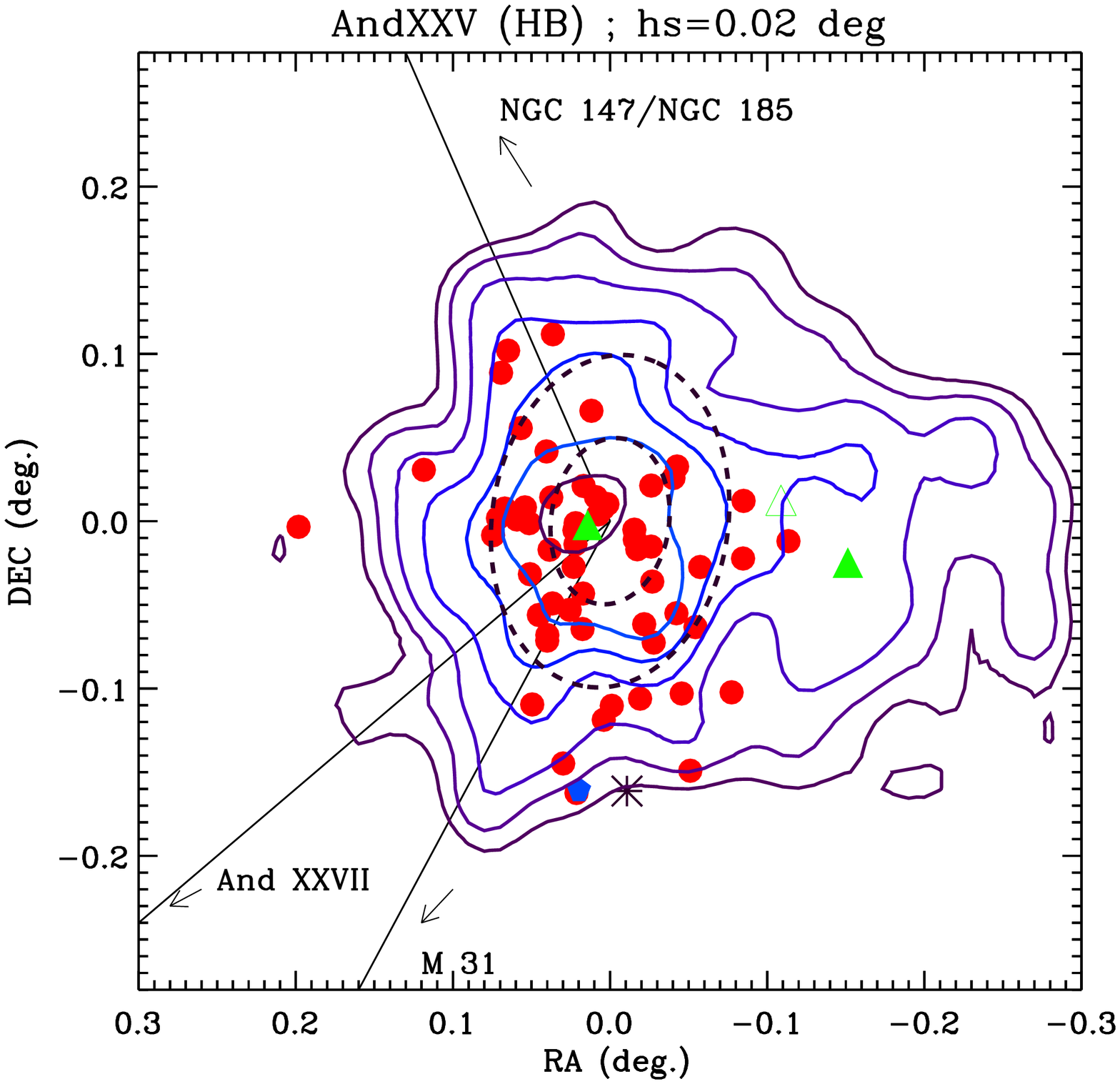}
\caption[]{$Left$: Isodensity contours of RGB stars in And~XXV. RA and DEC are the 
differential coordinates computed from the center of the galaxy given by Richardson et al. (2011).
The black dashed ellipses mark the areas within once and twice the r$_h$. The variable stars  
are shown with different colors and symbols (red circles for RR
Lyrae stars, green filled triangles for ACs (open green triangle for the uncertain AC/CC),  black asterisks for the ECL, blue pentagon for the unclassified variable). 
$Right$: Same as in the left panel but for And~XXV HB stars. 
}
\label{fig:isoden}
\end{figure*}

RGB stars were selected  from the CMD  in the right panel of Figure~\ref{fig:cmd},  showing stars contained in the whole LBT FOV. 
 The selection performed is marked with a magenta region in Figure~\ref{fig:cmd}.
The RGB isodensity contours are shown 
in the left panel of Figure \ref{fig:isoden} and reveal very interesting features 
of And~XXV structure. 

Moving outwards from  the galaxy center, the outermost contours appear to be elongated toward the position of  NGC147 and NGC185,
the major axis direction of the isodensity contours changes, and  finally points towards the direction of these two galaxies. 
Since isodensity twisting is mainly due to tidal effects \citep{dit1979,joh02}, we suggest that And~XXV is likely experiencing  tidal stirring from the relatively 
nearby couple NGC147/NGC185 on one side and from M31 in the orthogonal direction. Perhaps, And~XXV is infalling the M31 halo for
the first time and feeling both the tidal force of the NGC147/NGC185 couple and of the massive host galaxy M31.
Projected distances to NGC147/NGC185 and M31 are $\sim$95 kpc and $\sim$ 143 kpc, respectively, (adopting
distances reported  in \citealt{con12} for these systems). 
The right panel of Figure~\ref{fig:isoden}  shows the spatial distribution of the HB stars. This part of the CMD 
%
is heavily contaminated by background unresolved galaxies. Nevertheless it  is  still possible to see some of  features traced
by the RGB isodensities. A protrusion of HB stars extending in the direction of the M31 center highlights the tidal interaction between And~XXV and M31.


\section{GEP~I: AND~XXV's CENTER OR A NEW GLOBULAR CLUSTER ORBITING AND~XXV OR M31?}

\begin{figure*}[t!]
\centering
\includegraphics[trim=70 120 0 120 clip,width=6.9cm,height=6.6cm]{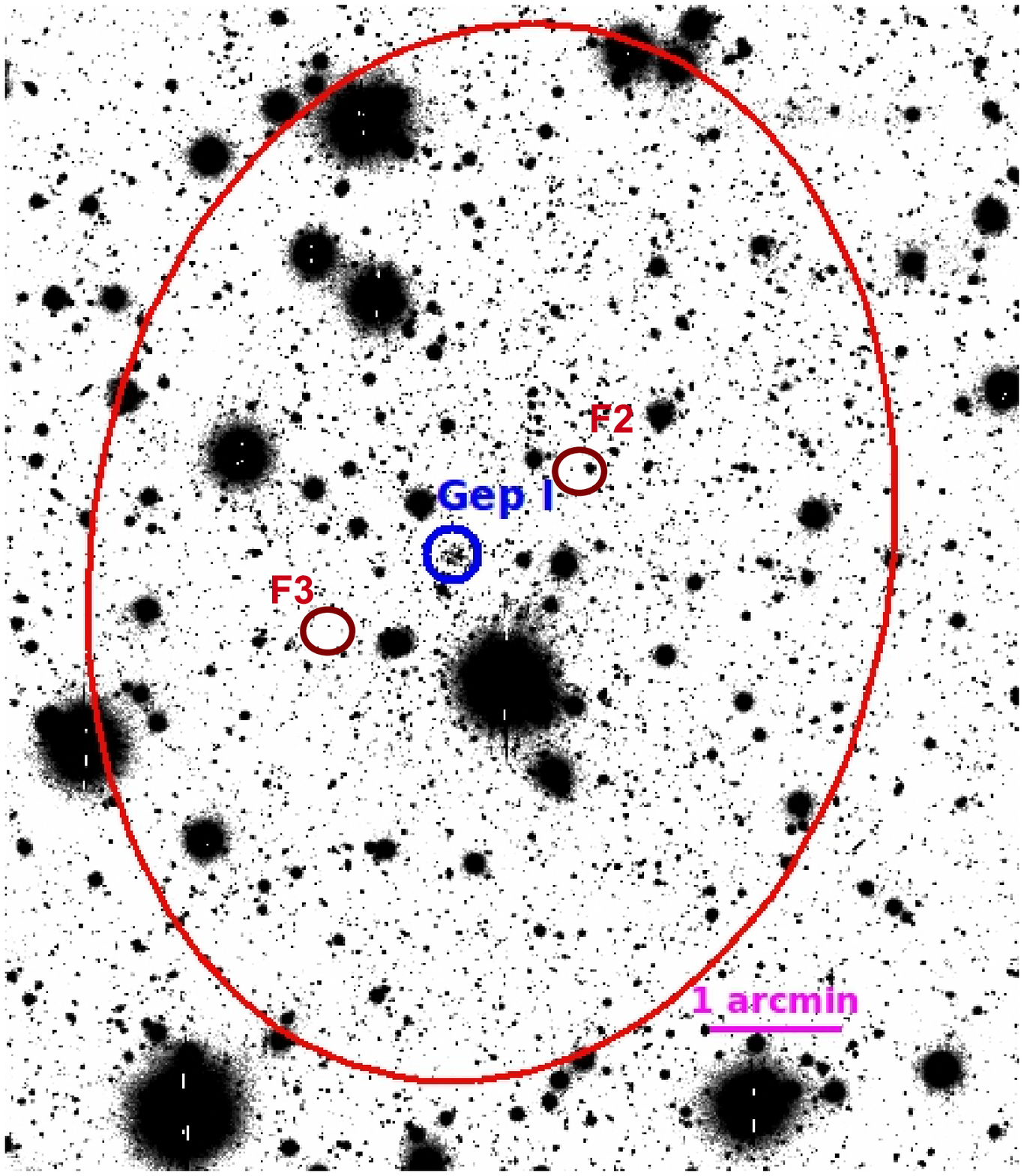}\includegraphics[width=6.9cm,height=6.6cm]{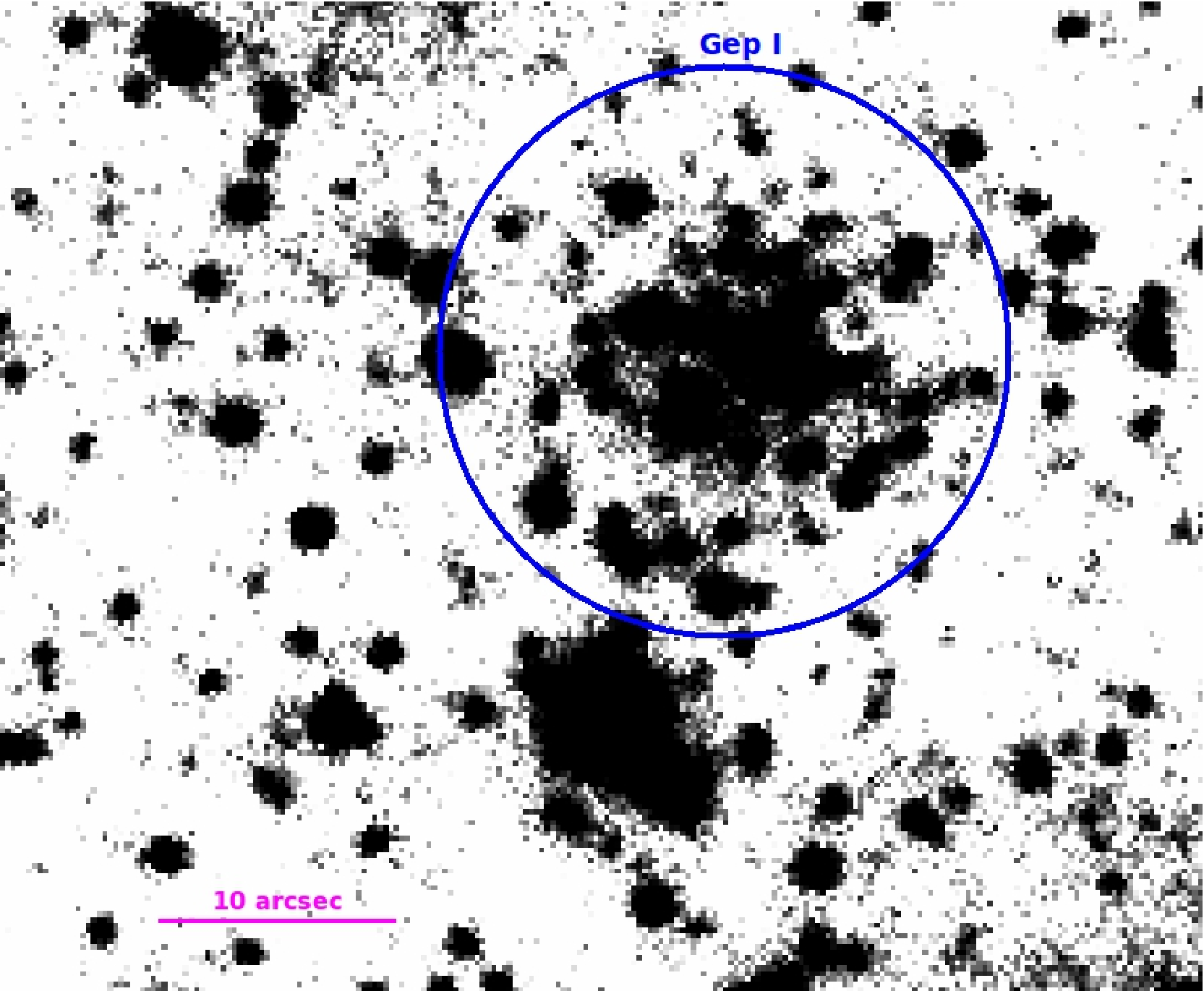}
\caption[]{$Left$:
Part of the deep image obtained by stacking all $B$-band frames. The position of Gep~I is highlighted with a blue circle. The positions of the comparison fields
 F2 and F3 are marked with red circles. 
The concentration of stars is near  the center of
And~XXV, the red ellipse shows the area delimited by once the r$_h$.  North is up and East to the left. $Right$: a closed in view of Gep~I.}
\label{fig:glob}
\end{figure*}

Visual inspection of the deep image obtained by stacking all the B-band LBC frames 
(for a total of 9.4 hours integration time) has revealed a spherical assembly of 
stars near And XXV's photometric center, that we named Gep~I\footnote{In the memory of a young colleague of us, Geppina Coppola, 
 prematurely passed away.}. This serendipity discovered concentration of stars  is centred around coordinates  
R.A.=00:30:10.579 dec.=+46:51:05.58 and is of $\sim$ 12 arcsec in diameter. 
A snapshot of Gep~I from our deep LBT B-band image is shown in Figure~\ref{fig:glob}. 
In the discovery paper \citet{rich11} note that And XXV falls close to the 2 arcmin gaps among 
CCDs in their observations (see Fig. 2 of \citealt{rich11}) significantly hampering the determination 
of And XXV’s centroid, for which they estimate a quite large $\sim$12 arcsec uncertainty. 
The difference between Gep~I and And~XXV center coordinates is $\sim$11 arcsec, hence, Gep~I is within 
the error box of  \citet{rich11} center coordinates of the galaxy.\\
The stellar concentration could either be 
And~XXV actual center or a new star cluster belonging to  And~XXV or M31.
Indeed, there is no known  Globular Cluster (GC) in the Revised Bologna Catalogue of the M31 GCs \citep{gal04}, nor any other 
known extended source within $\sim$20 arcsec of Gep~I center coordinates. 
Integrated photometry of Gep~I was performed using GAIA (Graphical Astronomy and Image Analysis Tool) starting from the center of the cluster 
and with circular apertures  ranging from  1 to  20 arcsec radius. With this method
we  estimated the half light 
radius of Gep~I to be $r_{h}\sim$ 6 arcsec and its integrated magnitudes, computed within twice the r$_h$,
are $B \sim$20.6 mag and $V\sim$20.0 mag, respectively. Assuming the distance modulus and reddening of And~XXV,  
we obtain absolute magnitudes of $M_{B}\sim -4.2$ mag and $M_{V}\sim -4.9$ mag, respectively. The $r_{h}$ of Gep~I at the distance 
of And~XXV corresponds to $\sim$25 pc in linear extension. 
Radius and absolute $V$ magnitude place Gep~I in the region of the $M_{V}-r_{h}$ plane that seems 
forbidden to ordinary GCs. Only the M31 Extended Clusters (ECs, \citealt{hux2011}) and the least luminous
among the MW Palomar GCs are found to lie in this region (see Figure 10 of \citealt{hux2011}) along with the nuclei 
of dwarf elliptical galaxies. This leaves open the possibility that Gep~I could indeed be the center of And~XXV.

\begin{figure}[t!]
\includegraphics[trim=40 25 0 0 clip, width=1.05\linewidth]{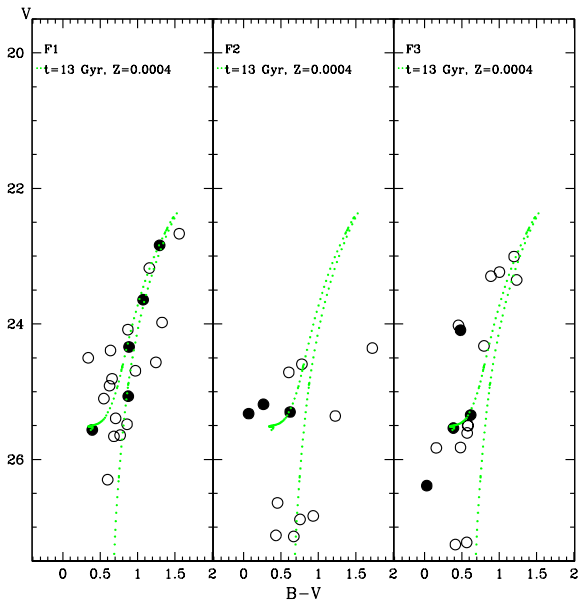}
\caption[]{
$Left$: CMD of sources in a circular area within a radius of 12 arcsec from the center of
Gep I (field F1). Filled circles are sources in our LBC photometric catalog selected using the
DAOPHOT $\chi$ and Sharpness parameters (-0.35 $\le$ Sharpness $\le$ 0.35 and $\chi < $ 1.5). The green
dashed line is an isochrone with t=13 Gyr and Z=0.0004 from Bressan et al. (2012). 
$Center$: same as in the left panel but for sources in F2. $Right$: same as in the left panel for
sources in F3.}
\label{fig:cmdgc}
\end{figure}

The stars that our photometry could resolve in this cluster are just a few and their position in the CMD 
is shown in the left panel of Figure~\ref{fig:cmdgc}. This CMD was done taking all the objects in a circular area of 12 arcsec 
in radius from the center of the stellar association (field F1). These stars are mostly red giants with only one or two HB stars. Their location
in the CMD is compatible with an old population placed at a Heliocentric distance of $\sim$750-800 kpc. 
Center and right panels of Figure~\ref{fig:cmdgc} show, for a comparison, the CMDs of sources in two circular regions F2 (central panel)
and F3 (right panel)  of a radius equal 
twice the estimated r$_h$ of Gep~I,
but centered $\sim$1 arcmin to the North-West and $\sim$ 1 arcmin to South-East with respect of the F1 centre, respectively.
 The positions of F2 and F3 in the LBC FOV are marked with red circles in Figure~\ref{fig:glob}.
The CMDs of fields F2 and F3 show no clear evidence of a single old stellar population similar to the one observed in field F1, 
thus confirming that in F1 there is a real concentration of old stars. Cleary, this ground-based CMD of Gep I is severely incomplete. 

To unveil the nature of Gep~I we have proposed follow up HST observations in order to resolve and characterize its stellar populations and precisely measure its distance. 
If the concentration of stars is proven to be a GC belonging to And~XXV the galaxy would be one of the few dSphs in the 
 Local Group known to host GCs, after Fornax and  Sagittarius.

Moreover, the position of this putative GC very close to the center of And~XXV (according to \citealt{rich11} coordinates), 
would resemble the case of M54 
which is  in the core of its parent galaxy, the Sagittarius dSph according to \citet{mon2005}.
Very recently a stellar cluster has been detected near the center of  the Eridanus~II  dwarf galaxy by \citet{crno2016}.
The authors claim that Eridanus II is the faintest galaxy to host a stellar cluster. In the same vein, 
And~XXV could be the faintest galaxy of the M31 complex to host a stellar cluster. 

\citet{zari2016} have suggested that most GCs may be hosted by undetected faint galaxies.
Our discovery in And~XXV along with \citet{crno2016}'s discovery of the cluster in Eridanus~II may lend support  to 
\citet{zari2016}'s claim and provide hints on the connection between GCs and dwarf galaxies.

\section{SUMMARY AND CONCLUSIONS}

Based on $B$, $V$ time series photometry obtained with LBCs at LBT
we have discovered 63 variable stars in the field of And~XXV dSph, of which 58 are RR Lyrae stars, 3 are ACs, 1 is an ECL 
and 1 is an unclassified variable. The average
period of the RRab stars classifies this galaxy as an Oosterhoff intermediate object.
The comparison of the galaxy CMD with isochrones  and the population of variable stars are compatible 
with a single old (9 -13 Gyr) and metal poor ([Fe/H]$\sim-1.8$ dex) stellar generation being the predominant component of And~XXV.
In the field of And~XXV we have detected a spherical concentration of stars 
near the galaxy center that could either be  a candidate globular cluster or the center of the galaxy.
And~XXV is the first of the M31 satellites we have investigated so far to lay on the  Great plane of Andromeda (GPoA) defined by \citet{iba2013}.
The stellar populations detected in And~XXV
differ somehow from the stellar populations we have identified in And~XIX and And~XXI, which both lay  
off the GPoA.
In particular, in And~XXV there is lack of an intermediate/young stellar population, compared to And~XIX
and And~XXI, and a very small number of ACs. 
Besides And~XXV, only  four other M31 satellites on the GpoA have been investigated so far to detect variable stars and have a metallicty [Fe/H]$\leq-1.5$ dex 
\citep[which is the condition to generate ACs, see e.g.][]{mar04},  namely:
And~I, And~III \citep{pri05}, And~XI and And~XIII \citep{yang12}.
\citet{pri05} found no ACs in And~I and two ACs in And~III (following our re-analysis in Section~\ref{sec:ac}), while in And~XI and And~XIII no ACs were detected. 
Although at present the statistic is still rather meager (5 out of 15 satellites laying on the plane) it seems that galaxies on the GPoA have no or very few ACs, hence, that they host no 
or very few intermediate-age stars\footnote{We note that And~XVI, a galaxy located at a projected distance of d$_{M31}$=323 kpc  from the M31 center, 
 but offset by only ~ 8 kpc from  the GPoA  \citep{paw2013}, also does not seem to contain ACs \citep{mon2016}.}.
 
\begin{figure}[t!]
\centering
\includegraphics[trim=40 15 0 0 clip, width=1.05\linewidth]{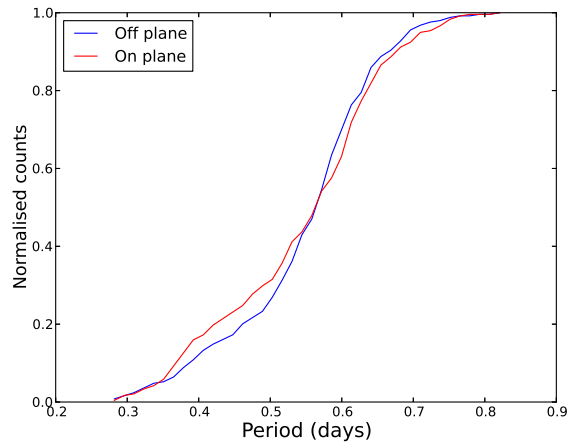}
\caption[]{Cumulative period distribution of the RR Lyrae stars in M31 satellite galaxies
on and off the GPoA.}
\label{fig:cumul}
\end{figure}

We have compared  the pulsation properties of the RR Lyrae population in the M31 dSph satellites 
 on and off the GPoA. 
The  M31 dSphs that have been studied for variability  are listed in the Table \ref{tab:rrl} along with their membership to the GPoA and the characteristics  
of their RR Lyrae populations. The average period of RRab stars for galaxies on and off the plane is  $\langle$P$_{\rm ab}\rangle$=$0.62\pm0.07$ d and 
$\langle$P$_{\rm ab}\rangle$=$0.60\pm0.06$ d, respectively. The fraction of RRc to the total number of 
RR Lyrae stars is f$_c$=0.31 and f$_c$=0.23 for on and off plane satellites.
Both populations are thus compatible with an Oo-Int classification, but with a slight tendency towards Oo-II type for the on plane satellites.
In Figure~\ref{fig:cumul} we plot the cumulative period distribution of the RR Lyrae stars in the two samples. A two-sample Kolmogorov-Smirnov (K-S) test was performed to  check if there are significant differences between the two populations. The p-value of the K-S test is p= 0.36 and at this
level of the investigation we cannot reject the hypothesis that the two RR Lyrae populations
have very similar pulsation properties. 
For the ACs distribution instead there is a clear difference considering that the total number of ACs  
for on plane galaxies is 3 while this number for the off plane galaxies is 21 (see Table \ref{tab:rrl} column 6). 
From the variable stars we can thus perhaps draw a first picture of the M31 satellite complex, in which all satellites had a common ancient 
star formation episode about 10-12 Gyr ago that led to the formation of the RR Lyrae stars. After that, the evolution history of on and off plane satellites  
started to differentiate and only the latter were able to retain enough gas to produce an intermediate-age stellar population that
also gave rise to significant numbers of ACs.

\begin{table*}
\caption[]{Properties of the variable stars in the Andromeda satellite galaxies}
\footnotesize
\label{tab:rrl}
\begin{center}
\begin{tabular}{l c c l l c c c}
\hline
\hline
\noalign{\smallskip}
Name    & N (RRab+RRc) &  $\langle$P$_{\rm ab}\rangle$  & f$_c$  & N (AC)&N (AC) confirmed*&  member & Reference\\
\hline
And~I   & 72+26       &    0.57   & 0.26 &    1?    &	0 & yes  &   (1)    \\ 
And~II  &   64+8      &   0.57    & 0.11 &    1     &	 0  &no  &   (2)    \\
And~III & 39+12       & 0.66      & 0.23 &    5?    &	 2 &yes  &   (1)     \\
And~VI  &   91+20     & 0.59      & 0.18 &    6     &	  4  &no  &   (3)	  \\
And~XI  &  10+5       &  0.62     & 0.33 &    0     &   0  &yes  &   (4)	 \\
And~XIII& 12+5        &  0.66     & 0.30 &    0     &	 0 &yes   &   (5)     \\
And~XVI &   3+6       &   0.64    & 0.33 &    0      &   0  & no$^1$    &  (5,6) \\
And~XIX & 23+8        &   0.62    & 0.26 &    8     &	  8 &no  &   (7)   \\	   
And~XXI &   37+4      &   0.63    & 0.10 &    9     &    9  &no  &   (8)  \\
And~XXV &   46+12     &   0.60    & 0.21 &    2	    &  1   &yes  &   (9) \\
\hline 
\end{tabular}
\end{center}
* see Section~\ref{sec:ac}\\
$^1$ offset by 8 Kpc from the GPoA \\
(1) \citet{pri05}; (2) \citet{pri04}; (3) \citet{pri02}; (4) \citet{yang12}; \\
(5) Mercurio et al. (2016, in preparation);(6) \citet{mon2016};(7) \citet{cus2013}; (8) \citet{cus2015};\\
(9) This work
\normalsize
\end{table*}

\acknowledgments

We warmly thank P. Montegriffo for the development and maintenance of the GRATIS software.
Financial support for this research was provided by  PRIN INAF 2010 (PI: G. Clementini) and by Premiale LBT 2013. 
The LBT is an international collaboration among institutions in the United States, Italy, and Germany. LBT Corporation partners 
are The University of Arizona on behalf of the Arizona university system; Istituto Nazionale di Astrofisica, Italy; 
LBT Beteiligungsgesellschaft, Germany, representing the Max-Planck Society, the Astrophysical Institute Potsdam, 
and Heidelberg University; The Ohio State University; and The Research Corporation, on behalf of The University of Notre Dame, University of Minnesota, 
and University of Virginia. We acknowledge the support from the LBT-Italian Coordination Facility for the execution of observations, 
data distribution, and reduction.
Facility: LBT

\end{document}